\documentclass[aps,prl,twocolumn,superscriptaddress]{revtex4}

\usepackage[normalem]{ulem}

\usepackage{amsmath}
\usepackage{amsfonts,amssymb,amsthm, bbm,braket}
\usepackage{graphicx}   
\usepackage{subfigure}  
\usepackage{amsbsy} 
\usepackage[bold]{hhtensor} 
\usepackage{natbib}
\usepackage{algpseudocode}

\usepackage{multirow}

\newcommand{\expect}{\mathbb{E}}

\newcommand\doi[1]{\href{http://dx.doi.org/#1}{doi:#1}}

\global\def \arxivmode {}

\ifx \arxivmode \undefined
\else
  
  \newcommand\arxivonly[1]{#1}
  \newcommand\prlonly[1]{}
\fi

\ifx \prlmode \undefined
\else
  
  \newcommand\arxivonly[1]{}
  \newcommand\prlonly[1]{#1}
\fi

\usepackage{hyperref} 
\usepackage{algorithm} 

\begin{document}

\title{Likelihood-free methods for quantum parameter estimation}

\author{Christopher Ferrie}
\affiliation{
Center for Quantum Information and Control,
University of New Mexico,
Albuquerque, New Mexico, 87131-0001}

\author{Christopher E. Granade}
\affiliation{
Institute for Quantum Computing and Department of Physics,
University of Waterloo,
Waterloo, Ontario, Canada}

\date{\today}


\begin{abstract}

In this Letter,  we strengthen and extend the connection between simulation and estimation to exploit simulation routines that do not exactly compute the probability of experimental data, known as the likelihood function.
Rather, we provide an explicit algorithm for estimating parameters of physical models given access to a simulator which is only capable of producing sample outcomes.
Since our algorithm does not require that a simulator be able to efficiently compute exact probabilities, it is able to exponentially outperform standard algorithms based on exact computation.
In this way, our algorithm opens the door for the application of new insights and resources to the problem of characterizing large quantum systems, which is exponentially intractable using standard simulation resources.
\end{abstract}


\maketitle

Much of physics is concerned with modeling complex behavior such that we can simulate systems of interest, and can infer properties of those systems.
On one hand, estimating the parameters of physical models given experimental data is critical to many practical objectives, such as precision metrology for frequency standards \cite{Udem2002Optical,Hinkley2013Atomic}, and to probing fundamental questions, such as gravitational wave detection \cite{Aasi2013Enhanced}.
On the other hand, by simulating physical models, we can understand properties of the systems that follow those models.
That is, by using simulation to reason about the probabilities of experimental data produced by physical models,
we can expose how experimental observations will depend on properties of interest.

Thus, these two concerns are not independent, such that parameter estimation can be broadly thought of as choosing as our estimated
model parameters those for which simulations predict the highest probability of obtaining data that
agrees with the observed experimental data.
Once we have estimated parameters for a model, we can use those parameters to predict the future
behavior of an experimental system by simulating according to those parameters.
In this way, simulation and statistical estimation are seen to be intimately related.

\begin{figure}[ht]\centering
   \includegraphics[width=.75\columnwidth]{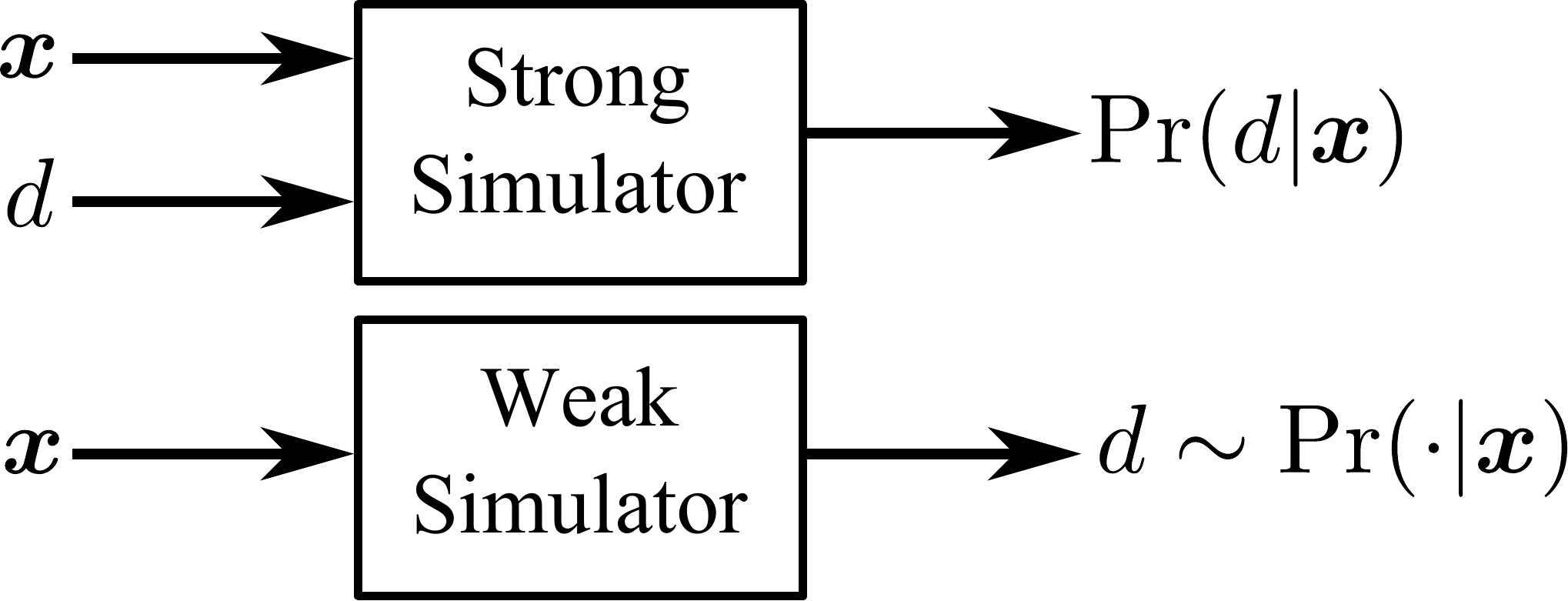}
  \caption{\label{fig:sims} Strong and weak simulators.  A strong simulator computes the value of the likelihood function $\Pr(\cdot|\vec{x})$, given a set of parameters $\vec{x}$ and data $d$.  By contrast, a weak simulator produces sample data $d$, drawn from the likelihood function, given only $\vec{x}$.}
\end{figure}

In this work, we present evidence of this relationship in the case of \emph{weak} simulation,
in which one has access only to samples from a simulator rather than the explicit distributions.
This is in contrast to a \emph{strong} simulator, which produces the exact probabilities of each possible outcome of an experiment (see Fig. \ref{fig:sims}).
The task of estimation is a statistical one and, in the language of statistics, strong simulation is equivalent to explicitly calculating the \emph{likelihood function}.
Many common estimation algorithms rely on explicit calculations of likelihood function and, hence, on strong simulation.
Here, we rectify the situation by providing a method to perform statistical estimation of parameters given access to only a weak simulator \footnote{We note that validation techniques (estimating ``closeness'' to some target as opposed to identifying an unknown set of parameters) have recently been proposed which take advantage of a similar notion of weak simulation \cite{Emerson2007Symmetrized, Knill2008Randomized, Magesan2011Scalable,Flammia2011Direct, daSilva2011Practical}}.  In addition to being generally applicable in the estimation of physical parameters, our approach is necessary in quantum certification protocols making use of quantum resources \cite{qhl1,qhl2}.

The distinction between strong and weak simulation is particularly important when considering quantum mechanical models, where we are only beginning to broadly appreciate the difference \cite{AA2011}.
In particular, it has been shown that many quantum mechanical models admit \emph{efficient} weak simulation on a classical computer where strong simulation is exponentially more difficult \cite{Jozsa2008Matchgates,VandenNest2010Classical,VandenNest2011Simulating,jv2013}.
Thus, a characterization method that
depends only on weak simulation can exhibit a large advantage over strong-simulation characterization methods.

This advantage is especially imperative in the case of quantum information,
as the number of parameters that must be measured in a tomographic experiment
grows exponentially with the number of qubits.
Though tomographic experiments have been carried out in
systems as large as several qubits \cite{Weinstein2002Experimental,Haffner2005Scalable},
the exponential nature of the problem prevents the extension of tomographic
methods to large-scale quantum information processing devices, such as those
currently being proposed \cite{jones_layered_2012,devoret_superconducting_2013}.
Thus, in order to develop useful quantum information processing devices,
it is necessary to develop novel and efficient statistical inference methods
that can exploit
prior information, reductions in model dimension and weak simulation.  Since our algorithm needs only a weak simulator, and does not require calculation of the likelihood function itself, we call our algorithm the \emph{likelihood-free parameter estimation} (LFPE) algorithm \footnote{We note that there exists a large class of ``likelihood--free'' methods, which has found use mostly in biological applications.  Our approach seems to be the closest to that in Ref. \cite{Sisson}.  However, contrary to our approach, all of the current likelihood--free methods available are based on rejection sampling.}.

Parameter estimation problems can be phrased in the following general terms.  To each physical model is associated a probability distribution $\Pr(d|\vec{x})$, where $d$ is the data obtained and where $\vec{x}$ is a vector parameterizing the system of interest.  In statistical parlance, this distribution is called the \emph{likelihood function}.

Now, suppose we have performed experiments and obtained a data set $D:=\{d_1,d_2,\ldots,d_N\}$.  We assume that experiments are statistically independent so that the likelihood function becomes
\begin{equation}
\Pr(D|\vec{x}) = \prod_{k=1}^N \Pr(d_k|\vec{x}).
\end{equation}
However, we are ultimately interested in $\Pr(\vec{x}|D)$, the probability distribution of the model parameters $\vec{x}$ given the experimental data.  We obtain this using use Bayes' rule:
\begin{equation}
\Pr(\vec{x}|D)=\frac{\Pr(D|\vec{x})\Pr(\vec{x})}{\Pr(D)},
\end{equation}
where $\Pr(\vec{x})$ is the \emph{prior}, which encodes any \emph{a priori} knowledge of the model parameters.  The final term $\Pr(D)$ can simply be found implicitly by normalizing the posterior.  Since each measurement is statistically independent given $\vec{x}$, the processing of the data can be done on- or off-line.  That is, we can sequentially update the probability distribution as the data arrive or post-process it afterward.   After all the data have been taken, we report the mean of the posterior distribution as our estimate of the parameters:
\begin{equation}\label{eq:expect}
\hat{\vec{x}}(D) = \mathbb E_{\vec{x}|D}[\vec{x}] =\int \vec{x} \Pr(\vec{x}|D) d\vec{x}.
\end{equation}
This method of parameter estimation is called Bayesian learning, and has been shown to be the optimal approach in a more general decision theoretic framework \cite{BlumeKohout2010Optimal}.  The meaning of this optimality is precisely that Eq.~\eqref{eq:expect} minimizes the mean squared error (MSE) figure of merit: $\text{MSE}(\hat{\vec{x}}) = \mathbb E_{\vec{x},D}[(\vec{x}-\hat{\vec{x}}(D))^2]$.

In order to efficiently compute the integral expectation in Eq.~\eqref{eq:expect}, we employ the \emph{sequential Monte Carlo} (SMC) method, which has been used for the purpose of Hamiltonian learning \cite{Granade2012Robust} and in the tomographic estimation of one and two qubit states \cite{Huszar2012Adaptive}, and in the continuous measurement of a qubit \cite{Chase2009Singleshot}.

The SMC method prescribes that we should approximate a distribution over model parameters with a distribution that has support only over a finite number of points (often referred to as \emph{particles}). Each particle is assigned a weight, informally thought of as its relative plausibility.  More concretely, we approximate the posterior distribution at the $N$th measurement by
\begin{equation}\label{eq:smc-posterior}
\Pr(\vec{x}|D) \approx \sum_{k=1}^n w_k(d_N) \delta(\vec{x} - \vec{x}_k),
\end{equation}
where the weights at each step are iteratively calculated from the previous step via
\begin{equation}\label{eq:weights}
w_k(d_{j+1}) = \Pr(d_{j+1}|\vec{x}_k)w_k(d_j) / \mathcal{N},
\end{equation}
where $\mathcal{N}$ is found implicitly by imposing the normalization condition $\sum_k w_k(d_{j+1}) = 1$.
The positions $\{\vec{x}_i\}$ of each particle are sampled according to the prior $\Pr(\vec{x})$.  The particle approximation can be made arbitrarily accurate by increasing the number of particles.  
The initial weights, when no data (denoted $d_0$) has been observed, are given by $w_k(d_0) = 1/n$ for all $k$.
This choice is made to ensure that the effective sample size $n_{\text{ess}} := 1 / \sum_i w_i^2$ is
initially $n$. As $n_{\text{ess}} \to 0$, the algorithm becomes numerically unstable and fails to explore the parameter space; this may be recovered by a resampling step \cite{LiuWest}. We explored some variants of this algorithm and presented it in much greater detail in reference \cite{Granade2012Robust}.
 
Equation \eqref{eq:weights} suggests that we require a full specification of the likelihood function $\Pr(d|\vec{x})$.   Suppose, however, we have access to only a weak simulator, which produces outcomes $d\sim \Pr(\cdot|\vec{x})$ \footnote{Note that we will overload the notation $\sim$.  Here we use $x\sim y$ to mean $x$ is random variable distributed according to $y$.  Later we use $x\sim y$ to mean $x$ is on the order of $y$ which is standard asymptotic notation formally defined to mean $x/y\to 1$.  The difference should be clear from context.}.  One extreme is to run the simulator many times and reconstruct $\Pr(d|\vec{x})$ from the simulated data---a meta-estimation problem.  At the other extreme is to perform estimation with only one sample per SMC particle.  The method truly becomes ``likelihood-free'' as we could not even hope to guess the functional form of the likelihood function from a single sample.  

In the extreme case where the weak simulator is used to very accurately compute the likelihood function via repeated sampling, the SMC algorithm does not change.  At the opposite extreme, when only a single sample is generated from the simulator per particle, we must modify the algorithm.  To this end, suppose we have obtained data $d$ from the experiment.  For each SMC particle, $\vec{x}_k$, we request a single sample $d'_k$ from our simulator and update the weight as follows:
\begin{equation}
w_k = \begin{cases} 1 &\text{ if }  d = d'_k\\
0 &\text{ otherwise}
\end{cases}.
\end{equation}

Between the two extremes of a single simulator sample per particle and enough to compute the likelihood function nearly exactly, we can \emph{approximatly} reconstruct the likelihood function sets of simulated data. In particular, for each datum $d$ and particle $\vec{x}_k$, we draw a \emph{set} of samples $D'_k$ from our simulator.  We then update the weights according to Eq. \eqref{eq:weights} with estimated likelihood function given by the naive maximum likelihood estimator
\begin{equation}
\Pr(d|\vec{x_k})\approx \frac{|\{d'\in D'_k: d' = d\}|}{|\{D'_k\}|} .
\end{equation}

As mentioned above, we will measure the performance of our algorithms with the mean squared error.  In our Bayesian setting, this is also the variance of the posterior distribution.  Since the experiments are assumed independent and identically distributed, the posterior variance will decrease as $O(1/N)$, where $N$ is the total number of measurements.  We appeal to standard Monte Carlo analyses which suggest that the SMC algorithm will increase this variance by at least $O(1/n)$, where $n$ is the number of SMC particles.  Now, if we use a weak simulator with a fixed experiment and particle number, the same statistical argument suggests that the variance will scale as $O(1/m)$, where $m$ is the number of simulator calls we use (per particle) to estimate the likelihood function.  Since the total number of samples is $nm$, we expect the mean square error to scale as
\begin{equation}
\operatorname{MSE}(\hat{\vec{x}}) \sim \frac{a}{N}  + \frac{b}{nm},
\end{equation}
for constants $a$ and $b$ depending only on the parameters of the problem.  

To verify these claims, we perform numerics.
Our example is that of a noisy photodetector where the efficiency of the photon source is $p$, which we would like to estimate.   This value is equivalent to the probability for the detector to click in the presence of no noise.  
In reality, \emph{dark counts} register clicks when no photon is present and \emph{losses} register no clicks when a photon is present.
Let these happen with probability $\alpha$ and $\beta$, respectively.
Then, given $p$, the probability for a click to actually happen is $\Pr(\text{click}|p) = p(1-\beta) + (1-p) \alpha$.
From these clicks, our task is to estimate $p$.

Aysmptotically, the posterior variance is given by the inverse of the Fisher information evaluated at, for example, the maximum likelihood estimate \cite{Clarke1999Asymptotic}.  The details of this calculation are presented in the appendix.
The result is the asymptotic bound
\begin{equation}\label{bound}
\operatorname{MSE}(p) \geq \frac{1}{6(1-\alpha-\beta)^2N},
\end{equation}
which we will use to verify our algorithm is near optimal.
In practice, $p$ will be a function of some parameters of interest,
$p=p(\vec{x})$. We restrict ourselves to this example in order to illustrate
the effects on inference due to weak simulation.

\begin{figure*}[ht]\centering
   \includegraphics[width=0.68\columnwidth]{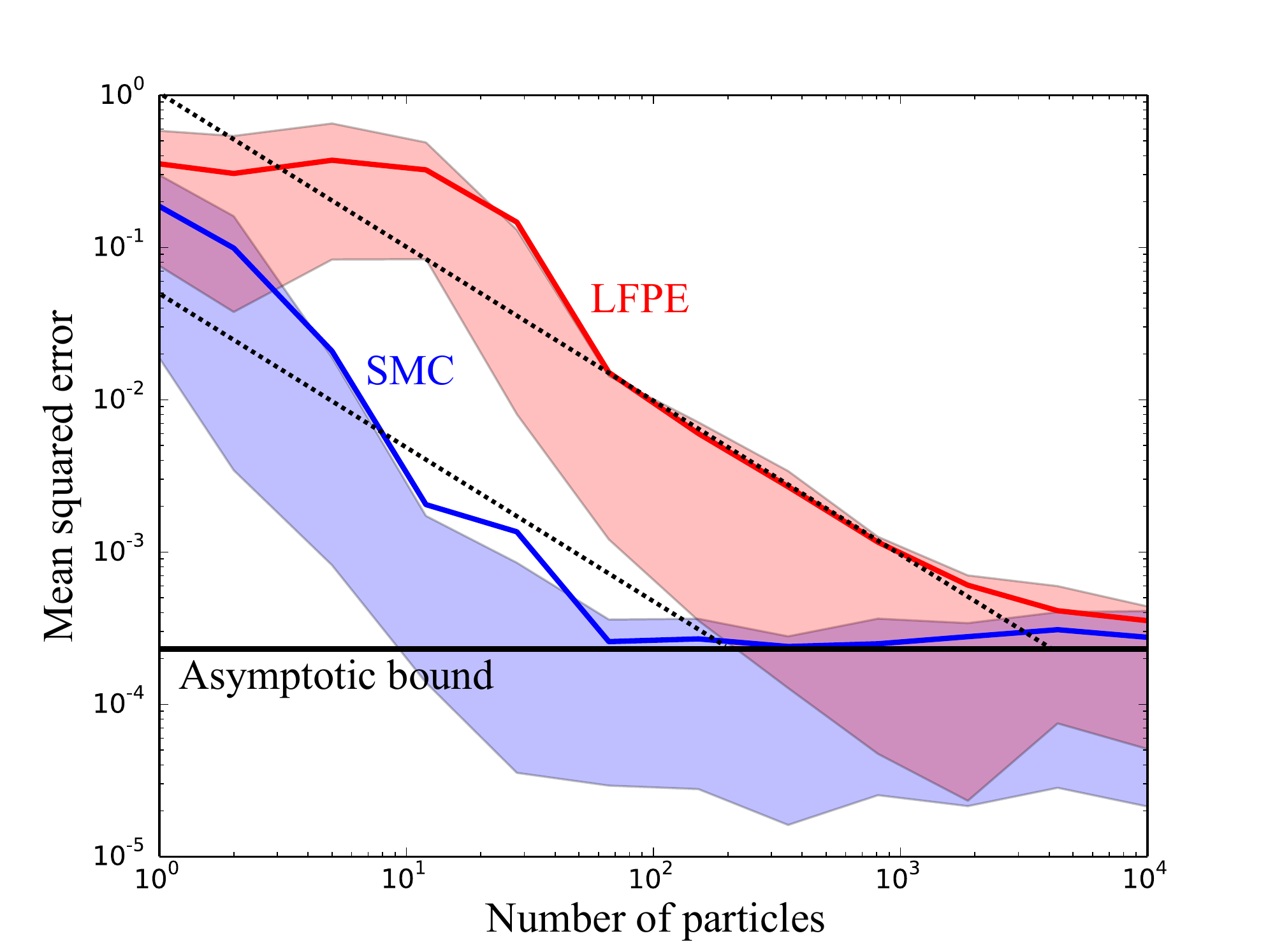}\includegraphics[width=.68\columnwidth]{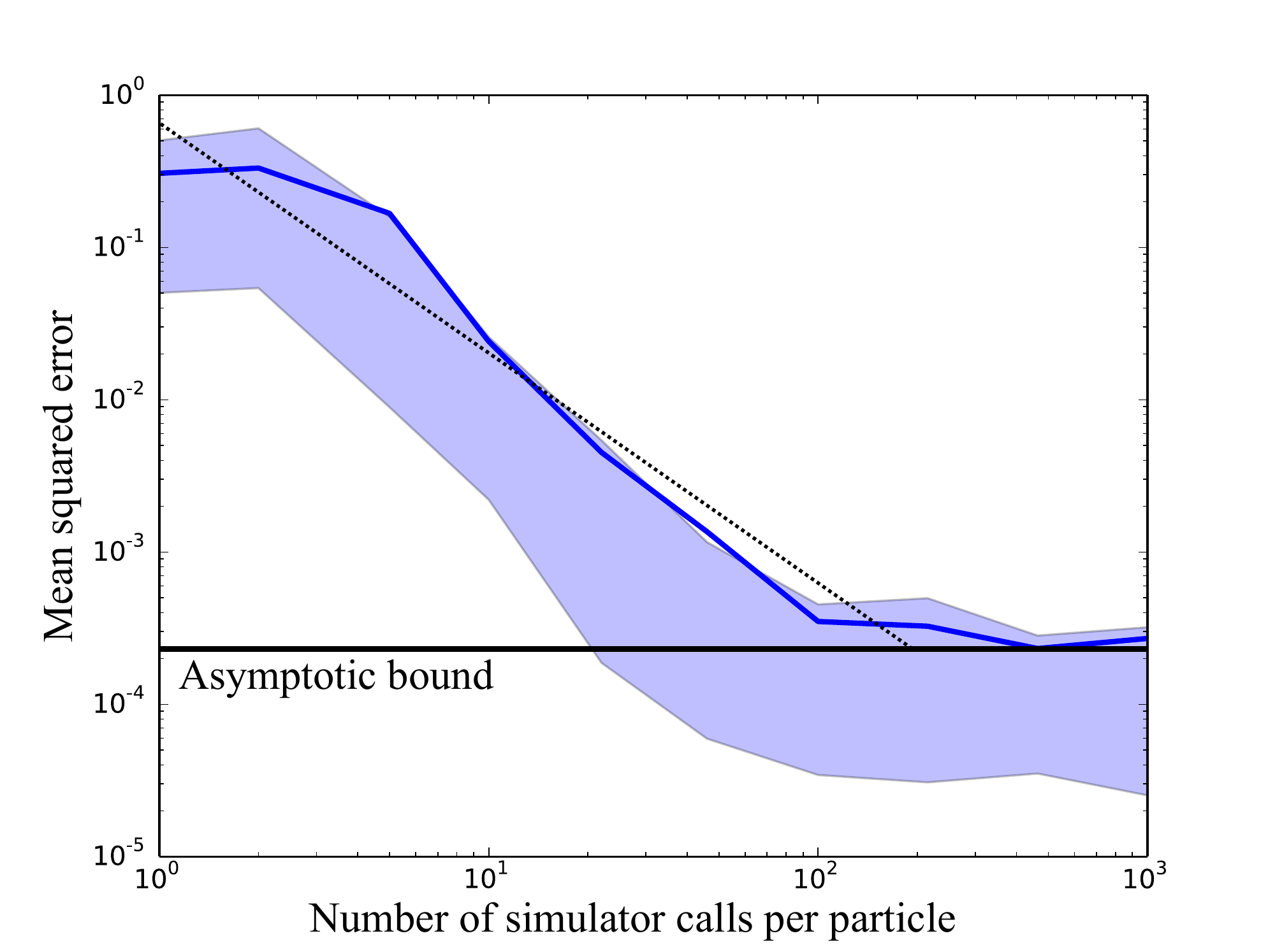} \includegraphics[width=.668\columnwidth]{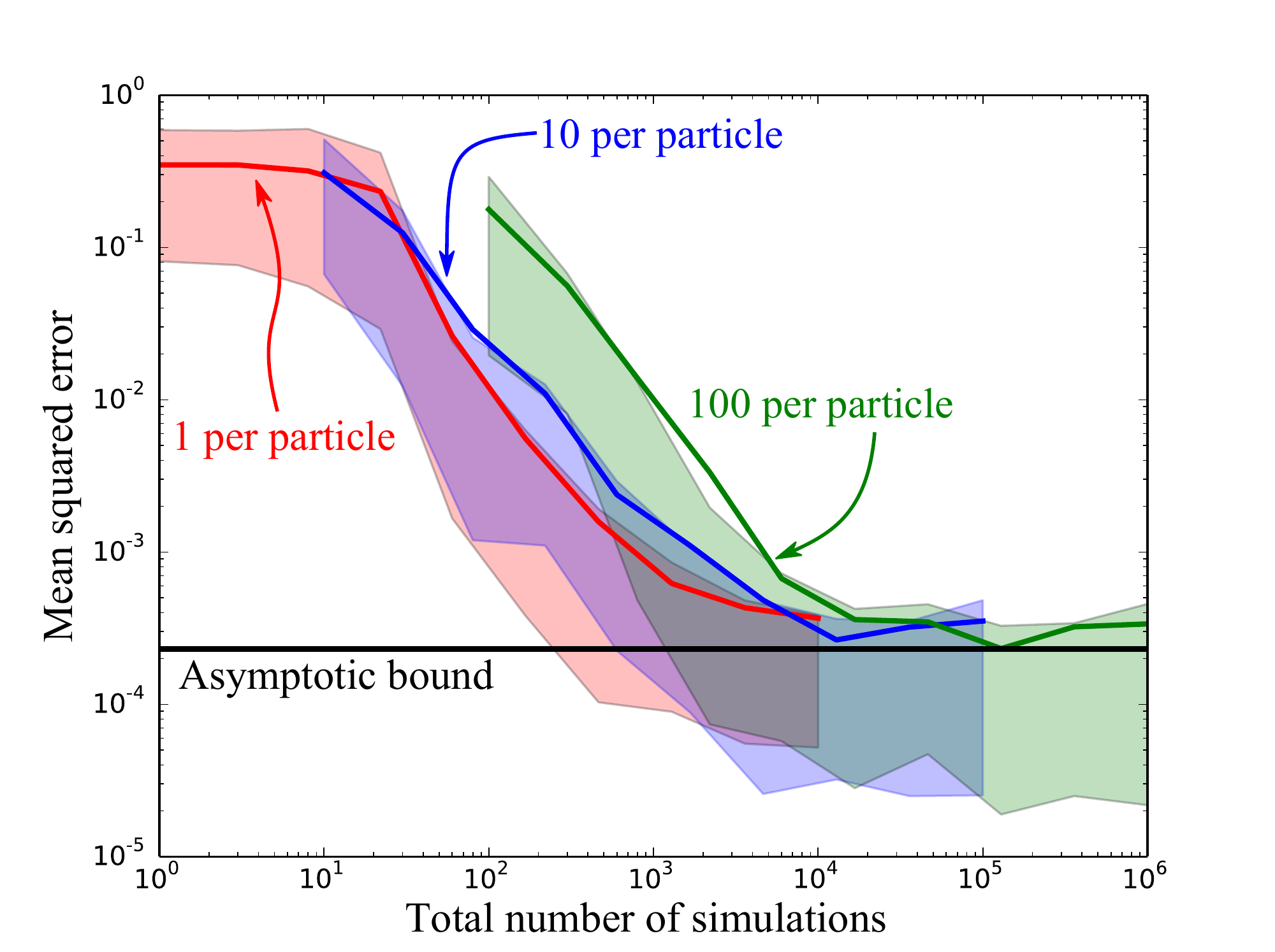}
  \caption{\label{fig:v_part} Left: the MSE using a strong simulator (SMC) and weak simulator (LFPE) as a function of the number of particles.  For the LFPE algorithm, a single sample from the simulator per particle is used to perform the inference.  The dotted line gives the conjectured $O(1/n)$ scaling.  Middle: the MSE using the LFPE algorithm as a function of the number of simulator calls per particles.  The number of particles is fixed at $n = 100$.  The dotted lines give the claimed $O(1/m)$ scaling.  Right: the MSE of the LFPE algorithm as a function of the \emph{total} number of simulator calls for varying number of simulator calls per particle.    Here, $\alpha=0.9$ and $\beta=0.05$ and the number of measurements is fixed at $N = 1000$.  The solid lines indicate the mean over 100 trials while the shaded areas represent the interquartile range (where the middle half of data lie).  The black solid line is the asymptotic bound given in Eq. \eqref{bound}.}
\end{figure*}

First, we verify that, given a fixed number of experiments, the MSE scales as $O(1/n)$ (where $n$ is again the number of SMC particles) for both the strong simulating SMC algorithm and likelihood-free weak simulation.  The data, plotted in Fig. \ref{fig:v_part} (left), bears out our expectations quite convincingly; even in the case of a \emph{single sample} from the simulator, the accuracy can be increased (at the expected $O(1/n)$ rate) until it reaches the bound given by Eq. \eqref{bound}. 
Next, in Fig. \ref{fig:v_part} (middle), we show that fixing the number of particles and varying the number of simulations per particle, $m$, results in an MSE that scales as $O(1/m)$.  Thus, as expected, the more accurately we can compute the likelihood function, the better our accuracy will be---but only up to a certain point.  That is, it is not advantageous to continue improving the accuracy of the estimate of the likelihood function beyond roughly $1/N + 1/n$ since the errors from finite particles and samples will begin to dominate.  

On the other hand, the strategy of estimating the likelihood via samples ignores the cost of simulation.  The total number of simulations is $nm$, the number of particles times the number of samples per particle.  In Fig. \ref{fig:v_part} (right), we plot the MSE against the \emph{total} number of simulator calls and find that, perhaps surprisingly, the likelihood--free approach of using a single simulator sample is best.  

In the above arguments, the total number of measurements was held fixed to verify the performance as a function of the algorithmic parameters.  If, on the other hand, simulations are relatively cheap compared to obtaining experimental samples, we would like to optimize performance by finding the appropriate number of simulated experiments without going beyond the redundancy noted above.  However, in most cases, the limit of accuracy is not \emph{a priori} known.  In such cases, we have devised an algorithm we call \emph{adaptive likelihood estimation} (ALE).  Essentially, our algorithm adaptively calls the simulator until we deem the accuracy in our estimate sufficient.  

For brevity, we will discuss the binary case with outcomes labeled $0$ and $1$.  The unknown probability $p_0: = \Pr(0 | \vec{x})$ can be treated as a parameter to be estimated.
In particular, since we have assumed that the data are conditionally independent given the model, repeatedly sampling the likelihood function will produce data that follows a binomial distribution with parameter
$p_0$. Estimating the parameter of a binomial distribution from sample data is a well-understood statistical problem.  Supposing $k$ $0$s were observed in $m$ trials, a typical estimator is
\begin{equation}\label{eq:addgamma}
\hat p_{0\gamma}(k) = \frac{k +\gamma}{m+2\gamma},
\end{equation}
where $\gamma$ is a free parameter.  These are called ``linear'' or ``add-$\gamma$'' estimators \cite{Lehmann1998Theory}.  The latter phrase is due to the equivalence to standard maximum likelihood estimation when adding $\gamma$ fictitious observations---also termed ``hedging'' \cite{BlumeKohout2010Hedged}.  These estimators can also be understood to arise from a Bayesian approach as well.  In particular, the estimator in Eq.~\eqref{eq:addgamma} is the posterior mean when using the following \emph{Beta distribution} as a prior \cite{Lehmann1998Theory}
\begin{equation}\label{eq:betadist}
\Pr(p_0) \propto p_0^{\gamma-1}(1-p_0)^{\gamma-1}.
\end{equation}
The posterior variance of this distribution can also be calculated as
\begin{equation}\label{eq:variance}
\hat{\sigma}_{0\gamma}^2(k) = \frac{(k+\gamma)(m-k+\gamma)}{(m+2\gamma)^2(m+2\gamma+1)} = \frac{\hat p_{0\gamma}(1-\hat p_{0\gamma})}{m+2\gamma+1}.
\end{equation}
Here we will use the value $\gamma=1$, as it corresponds to a uniform prior distribution.  We leave the optimization of this algorithmic parameter for future work. 

\begin{figure}[ht]\centering
   \includegraphics[width=.95\columnwidth]{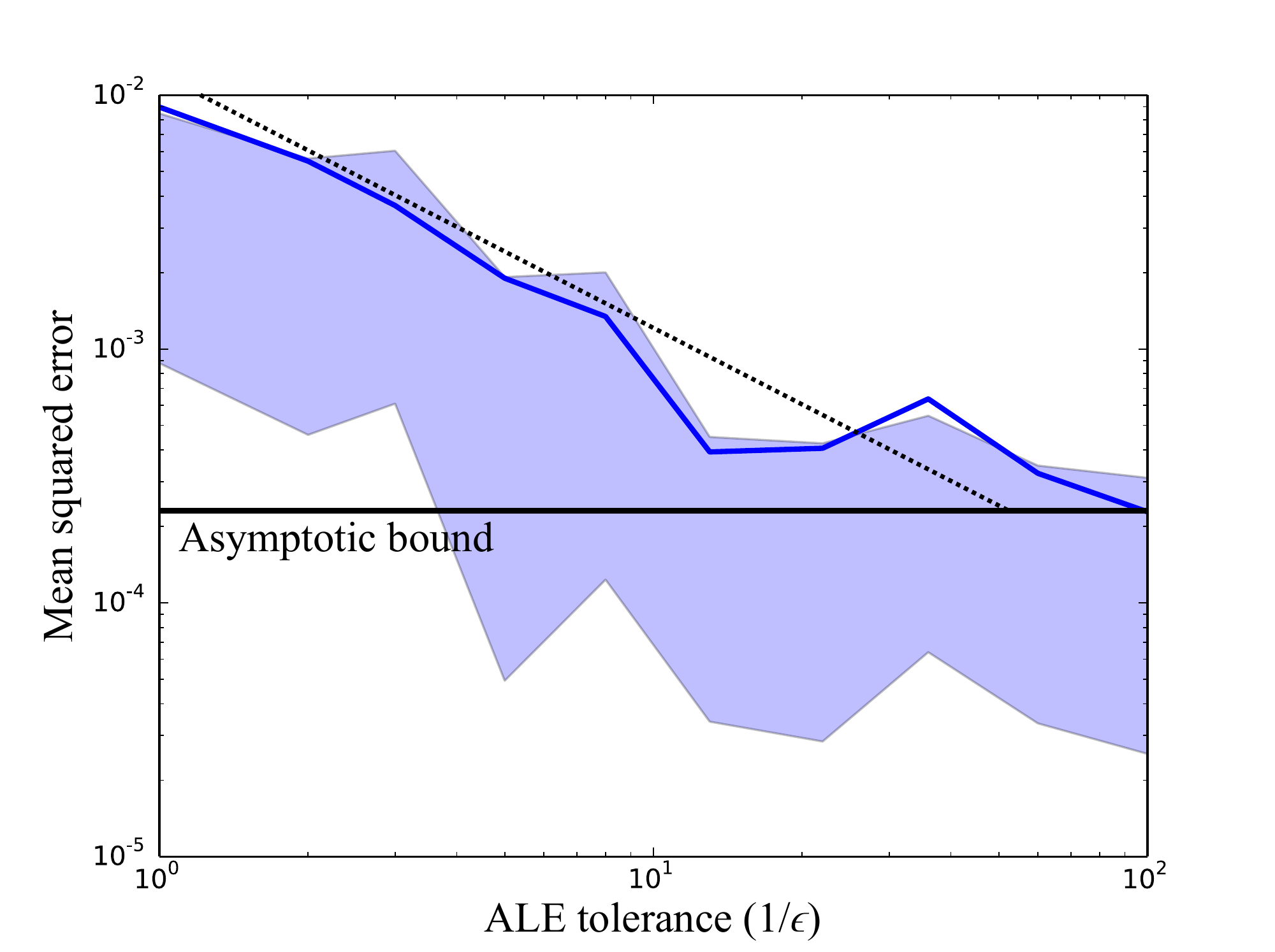}
  \caption{\label{fig:ale} The MSE using LFPE with adaptive likelihood estimation (ALE) as a function of the (inverse of the) ALE tolerance $\epsilon$.  As claimed, the MSE scales as $O(\epsilon)$ until it reaches the bound.  The parameters of the problem are as in Fig. \ref{fig:v_part}.}
\end{figure}

If we are willing to tolerate an error $\epsilon$ in our reconstruction of the likelihood, then we can check after each sample if $\hat{\sigma}_{0\gamma} < \epsilon$. If not, we collect more samples until the condition is met.  We therefore have a single quality parameter for this adaptive protocol: $\epsilon$.  Since this is our estimate of the variance in the estimate of the likelihood function, the MSE is expected to scale as $O(\epsilon)$ (for fixed measurement and particle number). Thus, as discussed above, the optimal choice will be $\epsilon \approx 1/N + 1/n$ since anything smaller will fast result in diminished returns---the MSE will be limited by either the number of measurements $N$ or particles $n$, depending on which is smaller.  We illustrate this with our example in Fig. \ref{fig:ale}.  

In this work, we have demonstrated an improvement of the sequential Monte Carlo parameter estimation
algorithm that allows for its extension to the case of weak (sampling) simulators.  For models with fast weak simulation available, our algorithm can be seen to provide dramatic
advantages in terms of classical computing costs
over sequential Monte Carlo alone and at minimal cost in estimation performance.  This extension allows for us to perform inference in subtheories of quantum mechanics that admit a large separation between
the tractibility of strong and weak simulation.

We have necessarily demonstrated these improvements for an example model in which the analytical solution was tractable.  In practice, if only weak simulation is available, then standard approaches to parameter estimation making use of calculations of the likelihood function do not apply and our method is necessary.  Within the confines of quantum theory, an ever growing class of weak simulation schemes have been proposed which have been proven to have an exponential separation in computational complexity between weak and strong simulation.  In addition to the large class of circuits identified by van den Nest \emph{et al}, others include simulating the evolution of  states with positive Wigner function \cite{Veitch2012Negative, Veitch2013Efficient, Mari2012Positive}.  In such cases, LFPE provides an exponential improvement in accuracy for a fixed amount of computational resources.  More recently, there have been proposals for the use of quantum resources (necessarily weak simulators) to aid in overcoming the complexity in simulating the physical model \cite{qhl1,qhl2}.  Such ideas could mitigate the need for classical simulators to certify near-future quantum devices which go beyond the classical regime, such as BosonSamplers \cite{AA2011}.  As the complexity of candidate quantum information processors grows, our algorithm provides a way forward to estimating properties of very large systems by exploiting the deep connection between simulation and estimation.

CF acknowledges funding from NSF grants PHY-1212445 and PHY-1005540 as well as NSERC of Canada. CG acknowledges support from the Canadian Excellence Research Chairs (CERC) program. The authors thank Josh Combes, D.G. Cory, Akimasa Miyake, Nathan Wiebe, Dan Puzzuoli and ``Referee B'' for helpful discussions and suggestions on improving the presentation.  


\appendix
\begin{widetext}
\section{Proof of Eq.~(9)}

First, we recall the example given in the main body of the paper, which is that of a noisy photodetector where the efficiency of the photon source is $p$, which we would like to estimate.   This value is equivalent to the probability for the detector to click in the presence of no noise.  
In reality, \emph{dark counts} register clicks when no photon is present and \emph{losses} register no clicks when a photon is present.
Let these happen with probability $\alpha$ and $\beta$, respectively.
Then, given $p$, the probability for a click to actually happen is $\Pr(\text{click}|p) = p(1-\beta) + (1-p) \alpha$.
From these clicks, our task is to estimate $p$.

Aysmptotically, the posterior variance is given by the inverse of the Fisher information evaluated at, for example, the maximum likelihood estimate \cite{Clarke1999Asymptotic}.  The maximum likelihood estimator if $k$ clicks are observed is 
\begin{equation}
\hat p_{\text{MLE}}(k) = \frac{k-N\alpha}{N(1- \alpha-\beta)}
\end{equation}
provided $N\alpha\leq k\leq N(1-\beta)$ (not a concern asymptotically).  The Fisher information of $N$ measurements, which is defined as
\begin{equation}
I(p) = \mathbb E_{k|p}\left[\left(\frac{d}{dp} \log \Pr(k|p) \right)^2 \right] ,
\end{equation}
can be calculated most simply in two steps given by the chain rule: $I(p) = I(q)(dq/dp)^2$. 
Since each measurement is a Bernoulli trial with probability $q=\Pr(\text{click}|p)$, the Fisher information for a single measurement is $I(q) = 1/q(1-q)$, yielding a Fisher information for $N$ measurements of
\begin{equation}
  I(q; N) = \frac{N}{q (1 - q)}.
\end{equation}
Taking the derivative and applying the chain rules yields
\begin{equation}
I(p) = \frac{(1-\alpha-\beta)^2 N}{(\alpha + (1-\alpha-\beta)p)(1-\alpha-(1-\alpha-\beta)p)}.
\end{equation}
With these facts we can say that asymptotically, the mean squared error in $p$ is
\begin{align}
\operatorname{MSE}(p) \sim \mathbb E_k[I(\hat p _{\text{MLE}}(k))^{-1}] \nonumber
    & = \expect_k \left[\left(
        \frac{(1-\alpha-\beta)^2 N}{(\alpha + (k/N) - \alpha)(1-\alpha-(k/N)+\alpha)}
    \right)^{-1}\right] \\
&= \mathbb E_k\left[\frac{k(N-k)}{(1-\alpha-\beta)^2N^3}\right] \nonumber \\
& = \frac{\mathbb E_k[k]N-\mathbb E_k[k^2]}{(1-\alpha-\beta)^2N^3},
\end{align}
where we have used that $(1 - \alpha - \beta) \hat{p}_{\text{MLE}} = (k/N) - \alpha$.

Letting the random variable $k$ be distributed according to a discrete uniform distribution on $\{0,\ldots,N\}$ and using known formulas for the first two moments $\mathbb{E}_k[k]$ and $\mathbb{E}_k[k^2]$ of the discrete uniform distribution, we finally arrive at
\begin{equation}
\operatorname{MSE}(p) \sim \frac1{6(1-\alpha-\beta)^2N}.
\end{equation}
In the finite $N$ regime, the variance will be larger, hence we obtain the bound
\begin{equation}
\operatorname{MSE}(p) \geq \frac{1}{6(1-\alpha-\beta)^2N},
\end{equation}
which is equivalent to that given in Eq.~\eqref{bound}.
\end{widetext}

\end{document}